\documentclass[
prd
,showpacs,amssymb,superscriptaddress,aps,
nofootinbib
]{revtex4-2}
\usepackage{graphicx}
\usepackage{color}
\input{epsf}

\usepackage{amsmath,amssymb}
\usepackage{bm}
\usepackage{times}
\usepackage{ulem}
\newcommand{\dalm}{\kern1pt\vbox{\hrule height 0.9pt\hbox{\vrule width 0.9pt
\hskip 2.5pt\vbox{\vskip 5.5pt}\hskip 3pt\vrule width 0.3pt}\hrule height 0.3pt}
\kern1pt}

\newcommand{\gsim}{\, \raisebox{-0.8ex}{$\stackrel{\textstyle >}{\sim}$ }}

\begin{document}

%\twocolumn[\hsize\textwidth\columnwidth\hsize\csname @twocolumnfals\endcsname

% For two column
%\wideabs{

\title{Gravitational wave asteroseismology of accreting neutron stars in a steady state}

\author{Hajime Sotani}
\email{sotani@yukawa.kyoto-u.ac.jp}
\affiliation{Astrophysical Big Bang Laboratory, RIKEN, Saitama 351-0198, Japan}
\affiliation{Interdisciplinary Theoretical \& Mathematical Science Program (iTHEMS), RIKEN, Saitama 351-0198, Japan}
\affiliation{Theoretical Astrophysics, IAAT, University of T\"{u}bingen, 72076 T\"{u}bingen, Germany}

\author{Akira Dohi}
\affiliation{Astrophysical Big Bang Laboratory, RIKEN, Saitama 351-0198, Japan}
\affiliation{Interdisciplinary Theoretical \& Mathematical Science Program (iTHEMS), RIKEN, Saitama 351-0198, Japan}

\date{\today}

% Abstract
\begin{abstract}
An accreting neutron star is potentially the gravitational wave source. In this study, we examine the gravitational wave frequencies from such an object in the steady state, adopting the Cowling approximation. We can derive the empirical relations independently of the mass accretion rate for the frequencies of the fundamental and 1st pressure modes multiplied by the stellar mass as a function of the stellar compactness, together with those for the 1st and 2nd gravity mode frequencies. So, once one simultaneously observes the fundamental (or 1st pressure) and gravity mode frequencies, one could constrain the neutron star mass and radius. In addition, we find that the luminosity can be well characterized by the mass accretion rate independently of the stellar mass and equation of state, if the direct Urca does not work inside the star. Since the luminosity from the neutron star with the direct Urca can deviate from this characterization, one could identify whether the direct Urca process works or not inside the star by observing the luminosity. Both information obtained from the gravitational waves and luminosity help us to understand the equation of state for neutron star matter. 
\end{abstract}

%
%%%%%%%%%%%%%%%%%%%%%%%%%%%%%%%%%%%%%%%%%%%%%%%%%
%  04.30.-w  :  Gravitational waves: theory
%  04.40.Dg :  Relativistic stars: structure, stability, and oscillations (see also 97.60.-s Late stages of stellar evolution) 
%  21.60.-n  :   Nuclear structure models and methods
%  21.65.Ef  :  Symmetry energy
%  26.60.+c  :  Nuclear matter aspects of neutron stars
%  26.60.Gj  :  Neutron star crust
%  97.10.Sj  :   Pulsations, oscillations, and stellar seismology 
%  97.60.Jd  :   Neutron stars (see also 26.60.+c Nuclear matter aspects of neutron stars in nuclear physics)
%%%%%%%%%%%%%%%%%%%%%%%%%%%%%%%%%%%%%%%%%%%%%%%%%
%]
% For two column
%}
\maketitle

%\baselineskip 24pt

%%%%%%%%%%%%%%%%%%%%%%%%%%%%%%%%%%%%%%%%%%%%%%%%
\section{Introduction}
\label{sec:I}
%%%%%%%%%%%%%%%%%%%%%%%%%%%%%%%%%%%%%%%%%%%%%%%%

Neutron stars born through core-collapse supernovae at the last moment of the massive star's life, are a unique object for probing physics under the extreme states, which are very difficult to realize on the Earth. For instance, the density inside the star easily exceeds the nuclear saturation density, $\rho_0\simeq 2.7\times 10^{14}$ g/cm$^3$, while the gravitational and magnetic fields around/inside the star become much stronger than those observed in the solar system \cite{ST83}. Therefore, the observations of neutron stars and/or their phenomena enable us to extract the signature of physics under such extreme states. 

In practice, the discoveries of massive neutron stars \cite{D10,A13,C20} can exclude the soft equations of state (EOSs), with which the expected maximum mass does not reach the observed mass. The detection of gravitational waves from the binary neutron star merger, GW170817 \cite{GW170817}, tells us the tidal deformability of the neutron stars, which leads to the constraint on the $1.4M_\odot$ neutron star radius, i.e., $R_{1.4}$ should be smaller than 13.6 km \cite{Annala18}. The Neutron Star Interior Composition Explorer Mission (NICER) equipped on the International Space Station makes constraints on the neutron star mass and radius, especially on compactness (ratio of the mass to radius), through the light bending due to the strong gravity induced by the neutron star, which is the relativistic effects for PSR J0030-0451 \cite{Riley19,Miller19} and PSR J0740-6620 \cite{Riley21,Miller21}. These astronomical observations essentially give us the constraint on more massive neutron star models than the canonical mass, which leads to a constraint on the EOS for a relatively higher-density region. On the other hand, the experimental constraints on the nuclear parameters are also important to constrain a low-mass neutron star model, which corresponds to the EOS for a relatively lower-density region, e.g., \cite{SNN22,SO22,SN23}.

The oscillation frequencies from the objects are another important astronomical information for extracting the stellar properties. Since the oscillation modes strongly reflect the conditions inside the star, one may see the interior properties by observing such oscillation frequencies as an inverse problem. This technique is known as asteroseismology, which is the same approach as seismology on Earth and helioseismology on the Sun. For example, using this technique, it is suggested that the crustal properties and/or neutron star models could be constrained by identifying the frequencies of the quasi-periodic oscillations observed in the magnetar flares with the crustal torsional oscillations, e.g., \cite{GNHL2011,SNIO2012,SIO2016,SKS23}. Once one observes gravitational wave signals from a neutron star, it must be helpful for us to understand the neutron star's mass, radius, and EOSs, e.g., \cite{ AK1996,AK1998,STM2001,SH2003,TL2005,SYMT2011,PA2012,DGKK2013,Sotani20b,Sotani21,SK21}. Moreover, this technique has been adopted in the supernova gravitational waves to understand the physical background, e.g., \cite{FMP2003,FKAO2015,ST2016,ST2020a,SKTK2017,MRBV2018,SKTK2019,TCPOF19,SS2019,ST2020,STT2021,SMT24}.

In a similar way, one may discuss the stellar properties including thermal effects, where the gravity ($g_i$-) mode oscillations are additionally excited together with the fundamental ($f$-) and pressure ($p_i$-) modes \cite{GA24}. In fact, the protoneutron stars born through the supernova explosion at the last moment of the massive stars' life, continuously cool down due to the neutrino emission until $\sim 10^5$ years and the photon emission after $\sim 10^5$ years. During the cooling phase, the thermal profile inside the neutron star gradually changes by keeping the beta equilibrium at each time step. Up to now, neutron star cooling has already been studied with various models by varying the EOS, neutron star mass, composition in the atmosphere, and the effects of nucleon superfluidity/superconductivity (see Refs.~\cite{YP04,Page06,Burgio2021} for the review). If the gravitational waves from these neutron stars are observed, one may extract the neutron star properties, which are difficult to constrain from the observation of the surface temperature and/or the luminosity \cite{KHA15,SD22}. Moreover, using a numerical code similar to the neutron star cooling, one can discuss the accreting neutron star system if one additionally takes into account the heating due to the accreting matter. The temperature inside the star changes due to mainly the neutrino cooling even with the accretion, which is common as in the (isolated) neutron star cooling without accretion, and the heating with accreting the matter. Then, it will eventually settle down to a steady state if thermal relaxation lasts a sufficiently long time. In this study, we focus on this phase of neutron stars and discuss to extract the stellar properties, using the technique of gravitational wave asteroseismology. In practice, the steady state of an accreting neutron star can become a continuous gravitational wave source induced by clumps of stochastically accreted matter~\cite{DM24}. We note that nuclear burning may happen on the neutron star surface, depending on the mass accretion rate and ignition conditions, which leads to the Type-I X-ray burst (for a review, see Ref.~\cite{2021ASSL..461..209G}). In this study, we neglect such a burst phenomenon, which will be discussed somewhere in the future.
Note that the properties of the Type-I X-ray burst are partially affected by the neutron star equation of state, mass~\cite{Dohi20,Dohi21,2023ApJ...950..110Z,2024ApJ...960...14D}, and cooling process~\cite{Dohi22}.

This manuscript is organized as follows. In Sec. \ref{sec:ANSs}, we show the models of accreting neutron stars, on which we will make linear analysis. In this study, as mentioned before, we focus only on their steady state. In Sec. \ref{sec:Oscillations}, we discuss the dependence of the gravitational wave frequencies on the properties of accreting neutron stars. Finally, we conclude this study in Sec. \ref{sec:Conclusion}. Unless otherwise mentioned, we adopt geometric units in the following, $c=G=1$, where $c$ denotes the speed of light, and the metric signature is $(-,+,+,+)$.

%%%%%%%%%%%%%%%%%%%%%%%%%%%%%%%%%%%%%%%%%%%%%%%%
\section{Steady state of accreting neutron stars}
\label{sec:ANSs}
%%%%%%%%%%%%%%%%%%%%%%%%%%%%%%%%%%%%%%%%%%%%%%%%

In this study, we focus on the steady state of accreting neutron stars. To construct such models, we consider the spherically symmetric neutron star at each time step, whose metric is given by
\begin{equation}
  ds^2 = -e^{2\Phi} dt^2 + e^{2\Lambda}dr^2 + r^2(d\theta^2+\sin^2\theta d\phi^2), \label{eq:metric}
\end{equation}
where $\Phi$ and $\Lambda$ are the metric functions depending on only $r$ at each time step, and $\Lambda$ is directly associated with the gravitational mass inside the position $r$, $m(t,r)$, as $e^{-2\Lambda}=1-2m/r$. Using this metric, the local luminosity, $L(t,r)$, and the local temperature, $T(t,r)$, are calculated with one dimensional general relativistic stellar evolutionary code \cite{Fuji84,Dohi20,Dohi24}, which is originally formulated in Ref.~\cite{Thorne77}. In concrete, the basic equations to be solved are Tolman-Oppenheimer-Volkoff equations and energy transport equations, latter of which is given as
\begin{gather}
   \frac{\partial(e^{2\Phi}L)}{\partial m_r}\bigg|_t = e^{2\Phi}\left({\cal E}_n + {\cal E}_g - {\cal E}_\nu\right), \label{eq:dL} \\
   \frac{\partial(\ln T)}{\partial (\ln P)}\bigg|_t = \min(\nabla_{\rm rad}, \nabla_{\rm ad}), \label{eq:TP} \\
   \frac{\partial m}{\partial m_r}\bigg|_t = \frac{\varepsilon}{\rho_0}\left(1-\frac{2m}{r}\right)^{1/2}, \label{eq:dmdmr}
\end{gather}
where $m_r(t,r)$ is the baryonic rest mass inside the position $r$ at time $t$; ${\cal E}_n$, ${\cal E}_g$, and ${\cal E}_\nu$ are the energy generation rates with crust heating, the gravitational energy release rate, and the neutrino emissivity (see below for detail); $P$, $\varepsilon$, and $\rho_0$ are the pressure, energy density, and rest mass density; and $\nabla_{\rm rad}$ and $\nabla_{\rm ad}$ are radiative and adiabatic temperature gradients, given by 
\begin{gather}
  \nabla_{\rm rad} = \frac{3\kappa PL}{64\pi\sigma_{\rm SB}T^4 m_r},  \label{eq:nabla_rad}  \\
  \nabla_{\rm ad} = \frac{\partial (\ln T)}{\partial (\ln P)}\bigg|_s.  \label{eq:nabla_ad}
\end{gather}
Here, $\kappa$ and $s$ are the opacity and specific entropy; and $\sigma_{\rm SB}$ denotes the Stefan-Boltzmann constant. Regarding the opacities, which generally depend on $\varepsilon$ and $T$, i.e., $\kappa = \kappa(\rho_0,T)$, we especially adopt the estimation of $\kappa$ for radiation~\cite{Schatz99}, for conduction with electrons and muons~\cite{Potekhin15}, and for conduction with neutrons~\cite{Baiko01}.

Meanwhile, ${\cal E}_n$ is the energy generation rate due to the accretion-induced exothermal nuclear reactions, such as the electron-capture, neutron emission, and pycno-nuclear fusions; ${\cal E}_g$ is the gravitational energy release rate; and ${\cal E}_\nu$ is the energy loss rate due to the neutrino emission. In practice, ${\cal E}_n$ is proportional to the mass accretion rate, $\dot{M}$, and $q_i$, where $q_i$ is the effective heat per nucleon on the $i$-th reaction surface. For density-dependence of $q_i$ values, we simply utilize the results obtained by Haensel \& Zdunik (2008) \cite{2008A&A...480..459H}, assuming the initial compositions of the nuclear-burning ashes ${}^{56}$Fe (see Table A.3 in Ref.~\cite{2008A&A...480..459H} for detail, and see also Ref.~\cite{2021PhRvD.103f3009L}).

Regarding ${\cal E}_\nu$ value, we take into account a slow cooling process, such as the modified Urca process; baryon bremsstrahlung; electron-ion bremsstrahlung; electron-positron pair creation; photo-neutrino process; and plasmon decay. We take the former two processes taken from \cite{2001PhR...354....1Y}, while we take the latter ones from \cite{1996ApJS..102..411I}\footnote{Note that we utilized older version of the neutrino emissivities for electron-ion bremsstrahlung; electron-positron pair creation; photo-neutrino process; and plasmon decay in \cite{SD22}.}.
%\hs{This time, can you use a different neutrino emissivities from that we used in \cite{SD22}? If so, the reference is enough to cite only \cite{1996ApJS..102..411I}?}
%\tg{Yes, emissivities in the crust are different from \cite{SD22}. Itoh et al. 1996~\cite{1996ApJS..102..411I} explicitly listed $\nu$ emissivities containing only $e,{}^A_ZX,\gamma$, therefore, citing this reference is enough (p.s. my cooling calculation since this time utilizes Itoh et al. 1996, which is applicable with a wide range of $(\rho,T)$). }
%Each energy loss rate is summarized in Ref.~\cite{Friman79}.
In addition to these slow cooling processes, we also consider the direct Urca process \cite{1981PhLB..106..255B,1991PhRvL..66.2701L} if it happens. Meanwhile, for simplicity, we do not consider the effects of superfluidity in this study, which may change the thermal profiles due to the pairing of baryons during the thermal evolution \cite{Potekhin19,Potekhin23}. Nevertheless, since we focus on only the steady state in this study, the suppression of baryon-specific heat, which is the main contribution as superfluid effects~(e.g., Refs. \cite{2009ApJ...698.1020B,2017PhRvC..95b5806C}), is not significant due to the balance of the heating and cooling rates in the steady state.
%\tg{ In the presence of the Direct Urca process, it is affected by the suppression of neutrino emissivities~\cite{}
%In TM1e EOS with $2.1~M_{\odot}$ stars, we consider two extreme cases where the DU process occurs or not since the temperature of superfluid models must lie between two models. However, I found that neutron superfluidity in the crust (suppression of specific heat) could affect thermal profiles. }

%\hs{In addition, the effect of superfluidity is also taken into account in the calculations? If so, can you put that effect? Even if not, can you put the expected effects?}
%\tg{No, some referee may point out them, since the actual accreting NSs must be in the superfluid states. A few referees (such as Alexander Y. Potekhin) would persist this, so I recommend you to designate him as an unsuggested referee (and as we are now fighting him in another paper of Noda et al. about our cooling code).}

Furthermore, ${\cal E}_g$ is given as follows. Using an Eulerian coordinate with $q$, which is the mass fraction of $m_r$ to the total gravitational mass, $M(t)=m(t,R)$, defined as
\begin{equation}
   q = \frac{m_r}{M(t)}, \label{eq:q}
\end{equation}
${\cal E}_g$ can be expressed with the non-homologous, ${\cal E}_g^{nh}$, and homologous, ${\cal E}_g^h$, components \cite{Fuji84}, i.e., ${\cal E}_g={\cal E}_g^{nh}+{\cal E}_g^h$, which are respectively given by 
\begin{gather}
   {\cal E}_g^{nh} = -e^{-\Phi}\left(T\frac{\partial s}{\partial t}\bigg|_q + \mu_i\frac{\partial N_i}{\partial t}\bigg|_q \right),  \label{eq:egnh} \\
   {\cal E}_g^{h} =  e^{-\Phi} \frac{\partial m}{\partial t}\bigg|_{m_r}
      \left(T\frac{\partial s}{\partial (\ln q)}\bigg|_t + \mu_i \frac{\partial N_i}{\partial (\ln q)}\bigg|_i \right), \label{eq:egh}
\end{gather}
where $\mu_i$ and $N_i$ respectively denote the chemical potential and number per total gravitational mass for the $i$-th elements. The total (gravitational) mass, $M(R)$, slightly depends on time (increases with time) due to the mass accretion. We note that the last term in ${\cal E}_g^{h}$ corresponds to the compressional heating induced by accretion. The details of how to calculate the thermal evolution in an accreting system, using the equations described above, are the same as in Refs.~\cite{Fujimoto84,Mat16,Dohi21}. 

\begin{figure}[tbp]
\begin{center}
\includegraphics[scale=0.6]{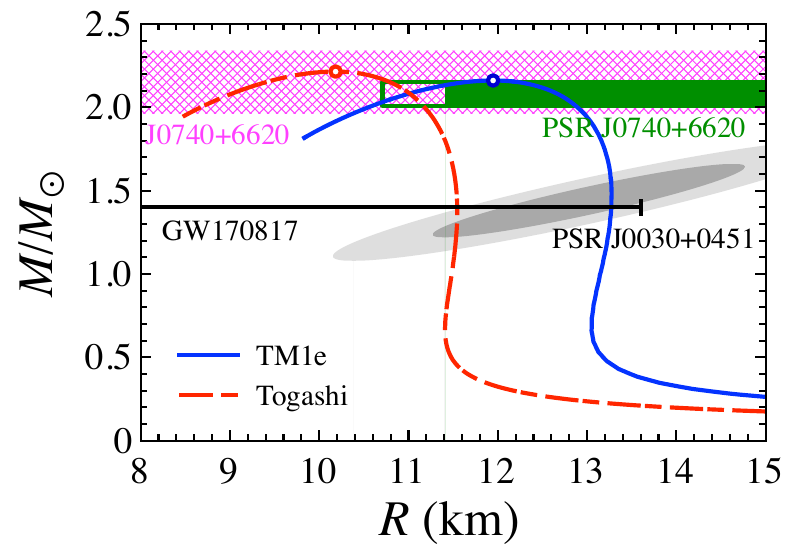} 
\end{center}
\caption{%%
Mass and radius relations for neutron star models constructed with TM1e (solid line) and Togashi (chain line). The circles denote the neutron star models with the maximum mass. For reference, we also show several constraints obtained from the astronomical observations. Namely, the $1.4M_\odot$ neutron star radius should be less than 13.6 km derived from the GW170817 event \cite{Annala18}, the mass of a massive neutron star, i.e., $M= 2.14^{+0.20}_{-0.18} M_\odot$ for PSR J0740+6620 \cite{C20}, the NICER constraint on the mass and radius for PSR J0030+0451 \cite{Riley19,Miller19}, and for PSR J0740+6620  \cite{Riley21,Miller21}. The inner and outer regions of the NICER constraint respectively denote the $1\sigma$ and $2\sigma$ levels \cite{Blaschke20}. 
}%%
\label{fig:MR}
\end{figure}

To solve this equation system, we especially consider two finite-temperature EOSs in this study. One is TM1e based on the relativistic mean-field framework \cite{Sumi19,Shen20}. Another is the Togashi constructed with the many-body variational method \cite{Togashi17}. In Table~\ref{tab:EOS}, we list the EOS parameters ($K_0$, $L$, and $\eta\equiv(K_0L^2)^{1/3}$ \cite{SIOO14}), the maximum mass of the spherically symmetric neutron stars, $M_{\rm max}/M_\odot$, and the critical mass above which the direct Urca process turns on, $M_{\rm DU}/M_\odot$. We note that $M_{\rm DU}$ strongly depends on $L$ or $\eta$ \cite{SD22}. Regarding the boundary between the neutron star crust and core, we fix that $P=10^{26.85}\simeq 7\times 10^{26}$ dyn cm$^{-2}$ as the crust surface \cite{Dohi21}. In Fig.~\ref{fig:MR}, we show the relation of neutron star mass and radius constructed with both EOSs together with several constraints from the astronomical observations, where one finds that neither of the two EOSs has been dismissed from current observations. From this figure, one may observe that TM1e and Togashi correspond to stiffer and softer EOSs allowed from the observations. In this study, we particularly consider the $1.4$, $1.8$, and $2.1M_\odot$ neutron star models constructed with Togashi and TM1e EOSs, when the neutron star reaches the steady state. We note that $2.1M_\odot$ neutron star model with TM1e EOS is above the critical mass for the direct Urca as shown in Table~\ref{tab:EOS}, i.e., the direct Urca naturally turns on inside this neutron star model. In addition to the model with the direct Urca, to see the neutron star mass dependence, we also consider the $2.1M_\odot$ neutron star model with TM1e EOS by artificially killing out the process of direct Urca. As mentioned before, we neglect the effects of superfluity in this study, but the temperature profile taken into account superfluidity must lie between two models with and without the direct Urca, e.g., Ref.~\cite{Dohi22}.

%%%%%%%%%%%%%%%%%%%%%%%%%%%%%%
%   TABLE 1
%%%%%%%%%%%%%%%%%%%%%%%%%%%%%%
\begin{table}
\caption{EOS parameters, $K_0$, $L$, and $\eta\equiv(K_0L^2)^{1/3}$, for the EOSs adopted in this study. In addition, the maximum mass of spherically symmetric neutron stars, $M_{\rm max}/M_\odot$, and the critical mass above which the direct Urca process turns on, $M_{\rm DU}/M_\odot$, are listed. We note that the direct Urca process does not occur with Togashi EOS even if the mass reaches the maximum mass. } 
\label{tab:EOS}
\begin {center}
\begin{tabular}{cccccc}
\hline\hline
EOS & $K_0$ (MeV) & $L$ (MeV) & $\eta$ (MeV) & $M_{\rm max}/M_\odot$ & $M_{\rm DU}/M_\odot$   \\
\hline
TM1e & 281 & 40.0  &  76.6 & 2.12 & 2.06  \\
Togashi & 245 & 38.7 & 71.6 & 2.21 & ---  \\
\hline \hline
\end{tabular}
\end {center}
\end{table}
%%%%%%%%%%%%%%%%%%%%%%%%%%%%%%

%%%%%%%%%%%%%%%%%%%%%%%%%%%%%%%%%%%
% Figure 1
%%%%%%%%%%%%%%%%%%%%%%%%%%%%%%%%%%%
%%%%%%%%%%%%%%%%%%%%%%%%%%%%%%%%%%%

\begin{figure}[tbp]
\begin{center}
\includegraphics[scale=0.52]{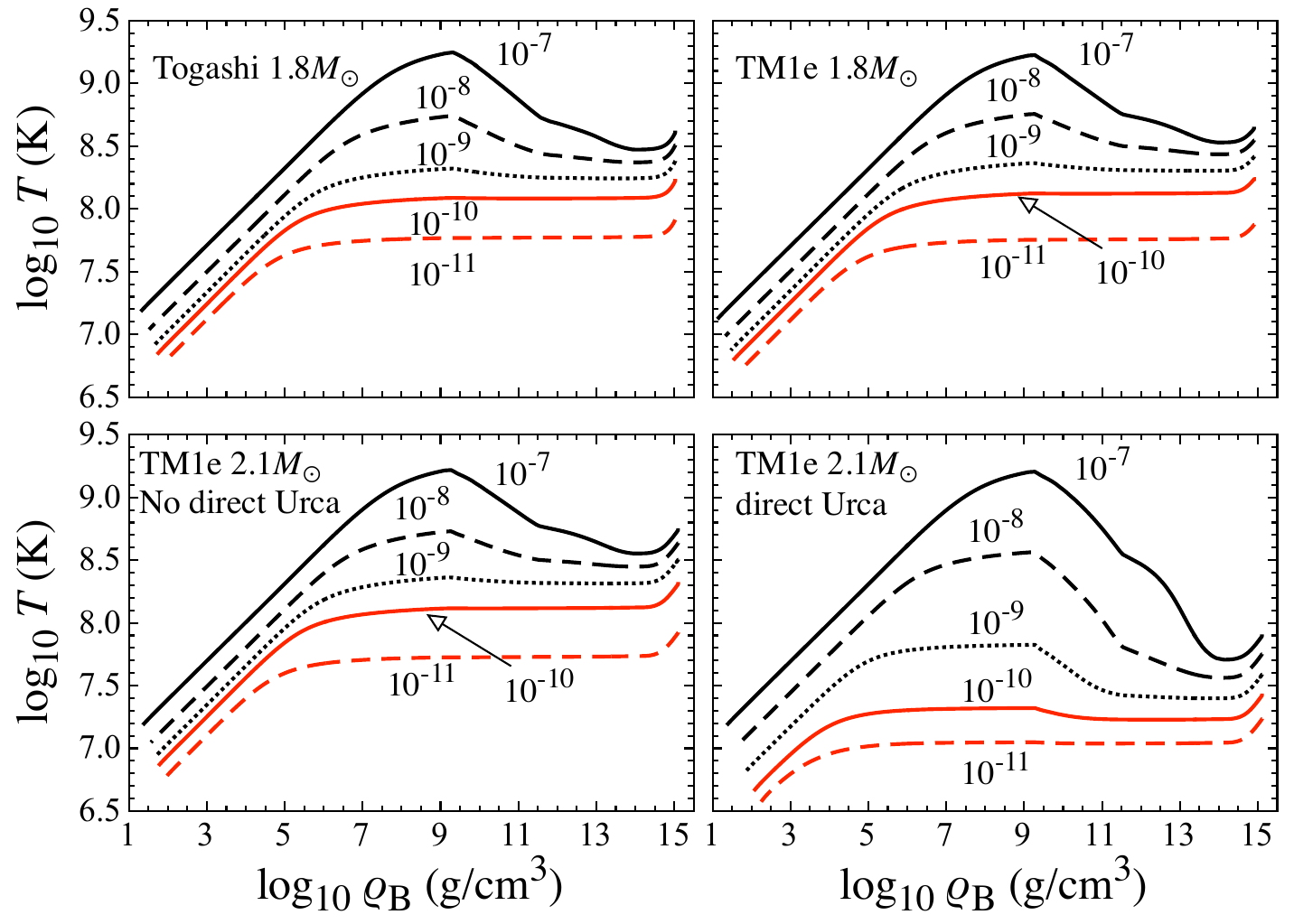} 
\end{center}
\caption{%%
The temperature profiles in the steady state of the accreting neutron star with various mass accreting rates. The thermal profiles are almost the same structure independently of the neutron star mass, unless the direct Urca turns on. The left-top and right-top panels correspond to the thermal profiles for the $1.8M_\odot$ neutron star models constructed with Togashi and TM1e EOSs, respectively, while the left-bottom and right-bottom panels are the results where the direct Urca neglects and turns on inside the star.
}%%
\label{fig:TrhoB}
\end{figure}

The thermal profiles inside the accreting neutron stars also depend strongly on $\dot{M}$. To obtain the steady state of the accreting neutron stars, first, we prepare the steady state with our fiducial value, $\dot{M}=4\times 10^{-9}M_\odot$/yr, as an initial model. Then, we change the accretion rate and calculate a long time evolution until the thermal profile becomes a steady state. 
The typical timescale in which the star becomes a steady state is $\sim 10^5-10^6$ yrs, e.g., \cite{Dohi21}, while the typical cooling timescale due to the neutrino emission in the neutron star is $\sim 10^4$ yrs \cite{Colpi01}.
In this study, we particularly consider the cases with $\dot{M}=10^{-11}$, $10^{-10}$, $10^{-9}$, $10^{-8}$, and $10^{-7}M_\odot$/yr, where the case with $\dot{M}=10^{-8}M_\odot$/yr roughly corresponds to the Eddington accretion rate. The X-ray burst happens, depending on the mass accretion rate, but we neglect such a bust phenomenon in this study, while we will discuss only the neutron star oscillations in the steady state. The resultant temperature profiles in the steady state are almost the same structure independently of the neutron star mass, unless the direct Urca turns on. In the left and middle panels of Fig.~\ref{fig:TrhoB}, we show the thermal profiles for $1.8M_\odot$ neutron star models constructed with Togashi and TM1e EOSs for various mass accretion rates. From this figure, one can observe that the temperature is enhanced at $\rho_{\rm B}\sim 10^9$ g/cm$^3$ for $\dot{M}\gsim 10^{-9}M_\odot$/yr due to the mass accretion. On the other hand, the right panel of Fig.~\ref{fig:TrhoB} corresponds to the thermal profile for $2.1M_\odot$ neutron star model with TM1e EOS, where the direct Urca process turns on inside the star. Owing to the direct Urca process, which is a fast cooling process, the neutron star core significantly cools down, which affects the profiles in the crust region.

%%%%%%%%%%%%%%%%%%%%%%%%%%%%%%%%%%%
% Figure 2
%%%%%%%%%%%%%%%%%%%%%%%%%%%%%%%%%%%
%%%%%%%%%%%%%%%%%%%%%%%%%%%%%%%%%%%

\begin{figure}[tbp]
\begin{center}
\includegraphics[scale=0.6]{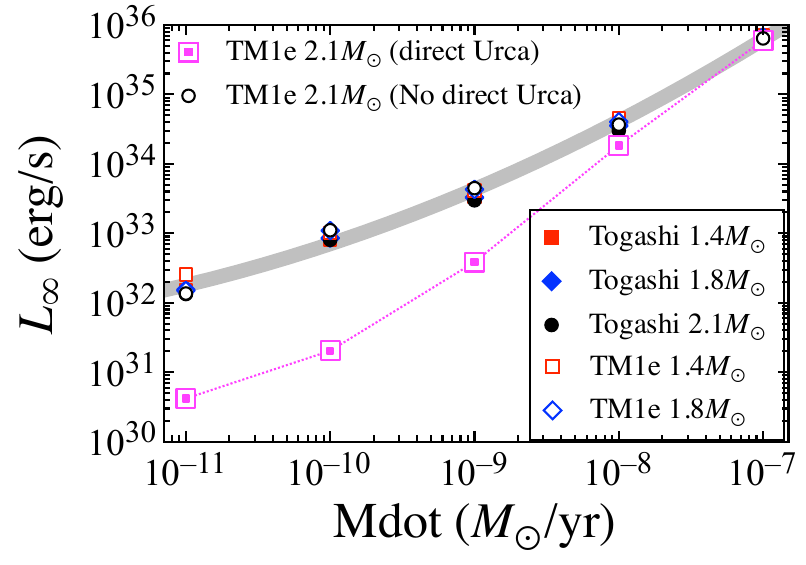} 
\end{center}
\caption{%%
The luminosity for the infinite observer, $L_\infty$, is shown as a function of the mass accretion rate, $\dot{M}$. The solid line denotes the fitting using the results for the stellar models, where the direct Urca process does not work, given by Eq.~(\ref{eq:Linf}). The double-squares are the results for the stellar models where the direct Urca process works, while the other marks are those for the stellar models where the direct Urca process does not work.
}%%
\label{fig:Linf}
\end{figure}

The thermal profile inside the star cannot directly be observed, while the luminosity radiated from the stellar surface is one of the important observables. The luminosity for an observer at infinity, $L_\infty$, is given by $L_\infty\equiv L(R)e^{2\Phi(R)}$ with the luminosity, $L(R)$, and potential, $\Phi(R)$, at the stellar surface. In Fig.~\ref{fig:Linf}, the values of $L_\infty$ in the steady state for various accreting neutron stars are shown as a function of $\dot{M}$. The double-squares are the results for the stellar model where the direct Urca process works, while the direct Urca process does not work in the other stellar models. This figure shows that $L_\infty$ hardly depends on the stellar mass and EOSs for neutron star matter if the direct Urca does not work. Using $L_\infty$ for the stellar models where the direct Urca does not work, we derive the fitting formula as a function of the mass accretion rate, such as
\begin{equation}
  \log_{10}L_\infty = 49.4518 + 2.6307 \left(\log_{10}\dot{M}\right) +0.097019\left(\log_{10}\dot{M}\right)^2, \label{eq:Linf}
\end{equation}
where $L_\infty$ and $\dot{M}$ are in the unit of erg/s and $M_\odot$/yr, respectively. This fitting is also shown in Fig.~\ref{fig:Linf} with the solid line. On the other hand, one can also observe that $L_\infty$ can significantly deviate from this fitting line if the direct Urca works inside the star. In practice, through the observation of $L_\infty$ in the steady state for the accreting neutron star, one could distinguish whether the direct Urca works or not inside the star, if $\dot{M}$ is less than $\sim 10^{-9}M_\odot$/yr. Once one can identify that the direct Urca process works inside the neutron star, it would give us the information between the stellar mass and EOS parameter through the relation between the critical mass above which the direct Urca works inside the neutron star and EOS parameter, e.g., Eq.~(9) or Eq.~(10) in Ref.~\cite{SD22}.

%%%%%%%%%%%%%%%%%%%%%%%%%%%%%%%%%%%
% Figure 3
%%%%%%%%%%%%%%%%%%%%%%%%%%%%%%%%%%%
%%%%%%%%%%%%%%%%%%%%%%%%%%%%%%%%%%%

%The crust EOS in the accreting neutron stars may be different from that for isolated neutron stars \cite{GC20}.

%%%%%%%%%%%%%%%%%%%%%%%%%%%%%%%%%%%%%%%%%%%%%%%%
\section{Asteroseismology}
\label{sec:Oscillations}
%%%%%%%%%%%%%%%%%%%%%%%%%%%%%%%%%%%%%%%%%%%%%%%%

Using the neutron star models provided in the previous section, we determine specific oscillation frequencies. In this study, we simply adopt the Cowling approximation to determine the frequencies. Namely, the metric perturbations are neglected (or the metric is fixed to background spacetime) during the fluid oscillations. The perturbation equations can be obtained by linearizing the energy-momentum conservation law. The concrete equations are completely the same as in Ref.~\cite{SKTK2019}. To integrate the perturbation equations, one has to impose the appropriate boundary conditions, i.e., the regularity at the center and the vanishing of the Lagrange perturbation of pressure at the stellar surface. In addition to these boundary conditions, one can adopt the normalization condition somewhere inside the star, owing to the nature of linear analysis. Then, the problem to solve becomes the eigenvalue problem with respect to the eigenvalue, $\omega$. The eigenfrequencies, $f$, are given as $f=\omega/(2\pi)$, using the resultant~$\omega$. We note that we set the neutron star surface, where the density becomes $10^6$ g/cm$^3$, for solving the eigenvalue problem. This is because the density sharply drops in the vicinity of the stellar surface, which leads to the difficulty of integrating the perturbation equations in the region where the density becomes very low.

Since we neglect the crust elasticity in this study, i.e., the matter is composed of a perfect fluid, one expects the excitation of the $f$-, $p_i$-, and $g_i$-modes as fluid oscillations \footnote{Once one takes into account the elasticity inside the star, the additional oscillation modes, such as the shear, interface, and torsional oscillations, can also be excited (e.g., \cite{Sotani23,Sotani24a}), which must tell us additional information, such as a crust elasticity.}
\footnote{The accretion might modify the crust elasticity, which leads to the change of the frequencies of the elastic oscillations. Even so, the frequencies of the fluid oscillations considered in this study hardly change, because they are less sensitive to the presence of the elasticity \cite{Sotani23,Kruger15}.}. 
One can generally identify the eigenmodes by counting the nodal number in the radial profile of the eigenfunction. That is, the $f$-mode has no node, while the $p_i$- and $g_i$-modes have $i$ nodes. The typical frequency of the $f$-mode becomes around a few kHz for a canonical neutron star model, where the frequencies of $p_i$-modes ($g_i$-modes) become higher (lower) than $f$-mode frequency. In addition, it is known that the $f$-mode frequency can be characterized by the stellar average density, $M/R^3$~\cite{AK1996,AK1998,STT2021}. Since the neutron star structures considered in this study are almost the same as those for cold neutron stars with the same mass, one expects that the $f$-mode frequencies even in the steady state of accreting neutron star must be similar to those for the cold neutron star with the same mass. The $f$-mode frequencies of the cold neutron stars constructed with Togashi EOS are 2.43, 2.55, and 2.65 kHz for $1.4M_\odot$, $1.8M_\odot$, and $2.1M_\odot$ stellar models, 
%2.426, 2.546, and 2.654 kHz for $1.4M_\odot$, $1.8M_\odot$, and $2.1M_\odot$ stellar models, 
we consider the eigenfrequencies, which are almost the same as those for the cold neutron stars, must correspond to the $f$-modes in the steady state of accreting neutron stars considered in this study.

\begin{figure}[tbp]
\begin{center}
\includegraphics[scale=0.6]{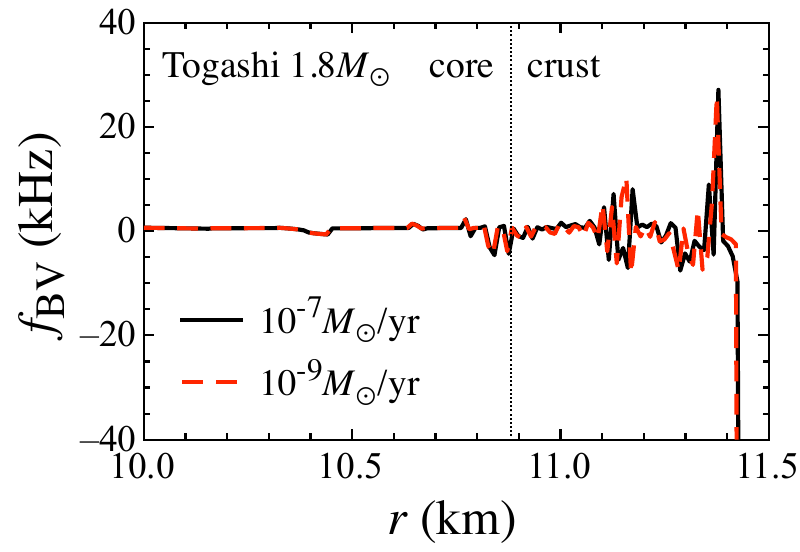} 
\end{center}
\caption{%%
The radial profile of the Brunt-V\"{a}is\"{a}l\"{a} frequencies for the $1.8M_\odot$ neutron star model constructed with Togashi EOS, where the solid and dashed lines denote the case with $\dot{M}=10^{-7}M_\odot$/yr and $10^{-9}M_\odot$/yr, respectively. For reference, we show the crust-core boundary with the vertical line. 
}%%
\label{fig:BV}
\end{figure}

However, we find that the behavior of eigenfunction seems to be strange in the vicinity of the stellar surface for some of the eigenmodes, which breaks the standard classification of the eigenmodes with the nodal number (see Appendix~\ref{sec:appendix_1}). This may come from the behavior of the Brunt-V\"{a}is\"{a}l\"{a} frequencies in the vicinity of the stellar surface is completely different from that for a protoneutron star produced through supernova explosion (e.g., see Fig. 4 in Ref.~\cite{ST2020a}). The Brunt-V\"{a}is\"{a}l\"{a} frequencies, which are associated with the convectional stability, are given by \footnote{Eq.~(2) in Ref.~\cite{ST2020a} is a typo. $2\pi$ in the denominator should be outside the square root as shown here.}
\begin{equation}
   f_{\rm BV} = {\rm sgn}({\cal N}^2)\sqrt{|{\cal N}^2|}/(2\pi),
\end{equation}
where ${\cal N}^2$ is locally determined through
\begin{equation}
   {\cal N}^2 = -e^{2\Phi-2\Lambda} \frac{\Phi'}{\varepsilon + p}\left(\varepsilon' - \frac{p'}{c_s^2}\right).
\end{equation}
In the above expression, the prime denotes the derivative with respect to $r$. With the Brunt-V\"{a}is\"{a}l\"{a} frequencies, one can discuss the local stability due to the convection, i.e., the region where $f_{\rm BV}<0$ is locally unstable. For example, the Brunt-V\"{a}is\"{a}l\"{a} frequencies for the $1.8M_\odot$ neutron star model constructed with Togashi EOS in the steady state with $\dot{M}=10^{-7}M_\odot$/yr and $10^{-9}M_\odot$/yr are shown with the solid and dashed lines in Fig.~\ref{fig:BV}. From this figure, one can observe that several unstable regions exist in the vicinity of the stellar surface, and the stellar surface becomes strongly unstable, where $f_{\rm BV}\simeq -200$ kHz. These local unstable regions may change the behavior of the eigenfunctions. We note that the neutron stars can globally oscillate with the eigenfrequencies unless $\omega^2$ becomes negative, even if the local unstable region exists inside the star, i.e., the star becomes unstable only if $\omega^2<0$.

Nevertheless, in the limit of zero accretion and zero temperature, the eigenfrequencies should approach those for a cold neutron star. So, in this study, we classify the eigenfrequencies, which are almost the same as the $f$-mode frequencies for a cold neutron star with the same mass, as the $f$-mode frequencies in the steady state of accreting neutron stars and the frequencies higher (lower) than the $f$-mode frequencies as the $p_i$-modes ($g_i$-modes) as usual.

%%%%%%%%%%%%%%%%%%%%%%%%%%%%%%%%%%%
% Figure 4
%%%%%%%%%%%%%%%%%%%%%%%%%%%%%%%%%%%
%%%%%%%%%%%%%%%%%%%%%%%%%%%%%%%%%%%

\begin{figure}[tbp]
\begin{center}
\includegraphics[scale=0.52]{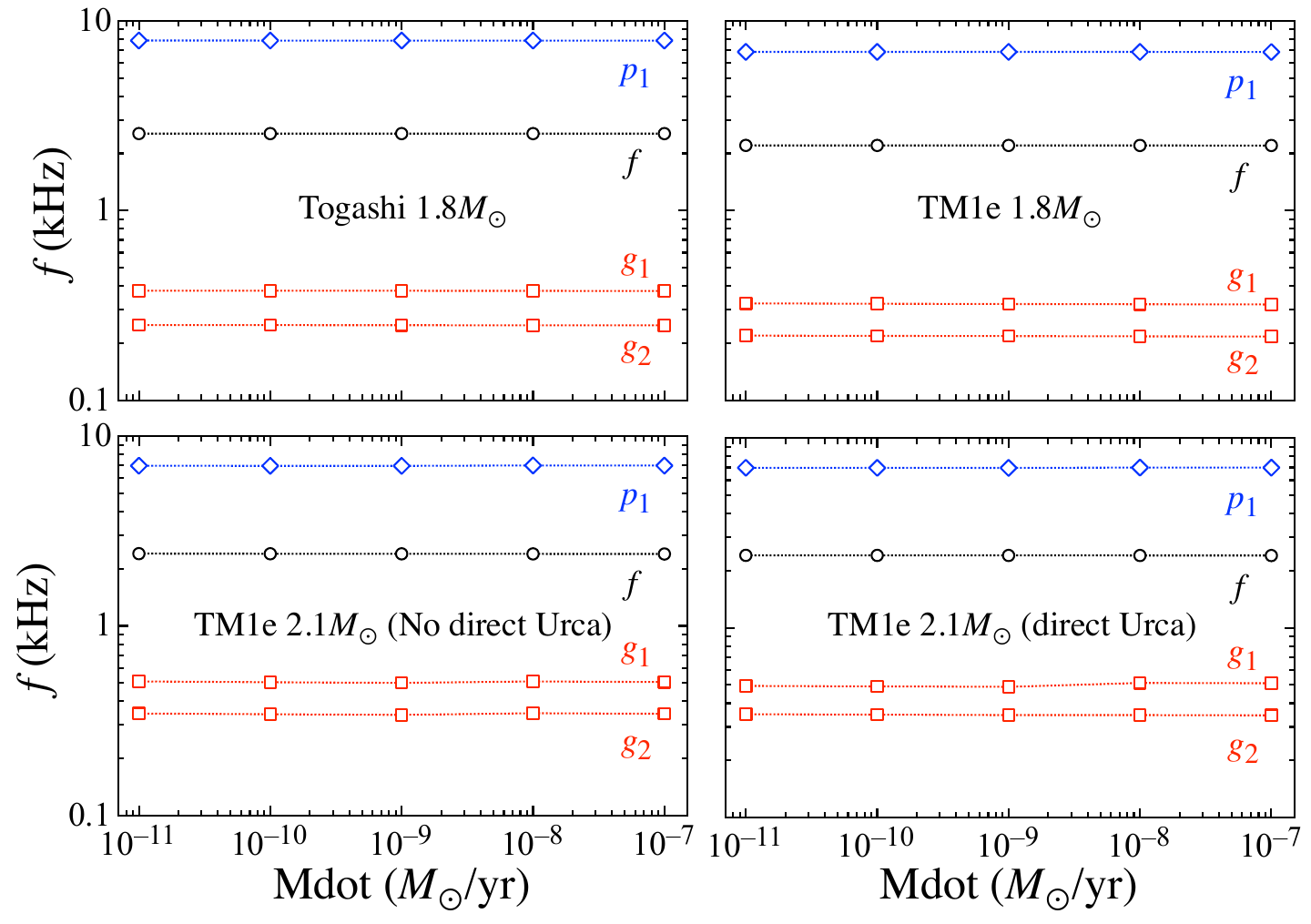} 
\end{center}
\caption{%%
For the neutron star models shown in Fig.~\ref{fig:TrhoB}, the excited frequencies of the $g_2$-, $g_1$-, $f$-, and $p_1$-modes are plotted as a function of the mass accretion rate. The frequencies depend on the stellar models, such as mass and EOS, but they are almost independent of the mass accretion rate.
}%%
\label{fig:fMdot}
\end{figure}

In Fig.~\ref{fig:fMdot}, we plot the excited frequencies on the neutron star models shown in Fig.~\ref{fig:TrhoB}, i.e., the $1.8M_\odot$ models constructed with Togashi and TM1e, where the direct Urca does not work, and $2.1M_\odot$ model constructed with TM1e, where we neglect the direct Urca and the direct Urca works, as a function of the mass accretion rate. From this figure, one can observe that the specific frequencies depend on the stellar models, such as the mass and EOS, while they hardly depend on the mass accretion rate. This behavior may be understood as follows. Since the $f$- and $p_i$-modes come from the acoustic oscillations of fluid elements, they (especially the $f$-mode frequencies) basically depend on the stellar mass and radius, i.e., they weakly depend on the thermal effects by nature. On the other hand, the $g_i$-modes can be excited due to the thermal effects characterizing with the Brunt-V\"{a}is\"{a}l\"{a} frequencies, $f_{\rm BV}$, but the radial profile of $f_{\rm BV}$ inside the neutron star core are almost independent of the mass accretion rate and it depends only in the crust, as shown in Fig.~\ref{fig:BV}. The deviation of $f_{\rm BV}$ inside the crust may make it difficult to control the properties of the global $g_i$-mode oscillations, because the crust thickness is only less than $\sim 10\%$ of the stellar radius and the density is much lower than that in the core. Nevertheless, one can see a tiny deviation of the $g_1$-mode frequencies with $\dot{M}=10^{-9}$ and $10^{-8} M_\odot$/yr for the $2.1M_\odot$ neutron star model with direct Urca constructed with TM1e. This may come from the fact that the thermal mountain (the local maximum in the thermal profiles) inside the crust with $\dot{M}=10^{-8}$ and $10^{-7} M_\odot$/yr is much larger than that in the other models, owing to the direct Urca process (see Fig.~\ref{fig:TrhoB}).

%%%%%%%%%%%%%%%%%%%%%%%%%%%%%%%%%%%
% Figure 5
%%%%%%%%%%%%%%%%%%%%%%%%%%%%%%%%%%%
%%%%%%%%%%%%%%%%%%%%%%%%%%%%%%%%%%%

\begin{figure}[tbp]
\begin{center}
\includegraphics[scale=0.5]{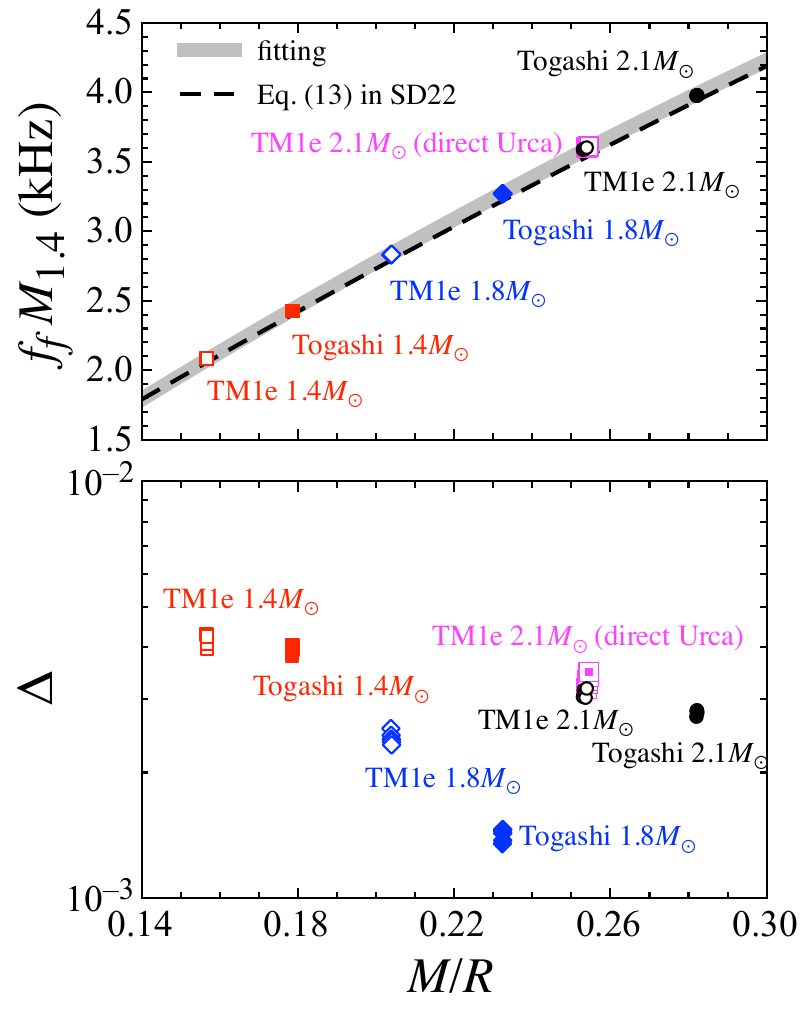} 
\end{center}
\caption{%%
The $f$-mode frequencies multiplied with $M_{1.4}\equiv M/(1.4M_\odot)$ are shown as a function of $M/R$ for various accreting neutron star models in the steady state. The solid line denotes the fitting given by Eq.~(\ref{eq:ffM}), while the dashed line denotes the fitting in the neutron star cooling derived in Ref.~\cite{SD22}. The bottom panel denotes the relative deviation of the $f$-mode frequencies determined with the eigenvalue problem from the expectation with the fitting given by Eq.~(\ref{eq:ffM}), calculated with Eq.~(\ref{eq:Delta}).
}%%
\label{fig:ffM}
\end{figure}

\begin{figure}[tbp]
\begin{center}
\includegraphics[scale=0.5]{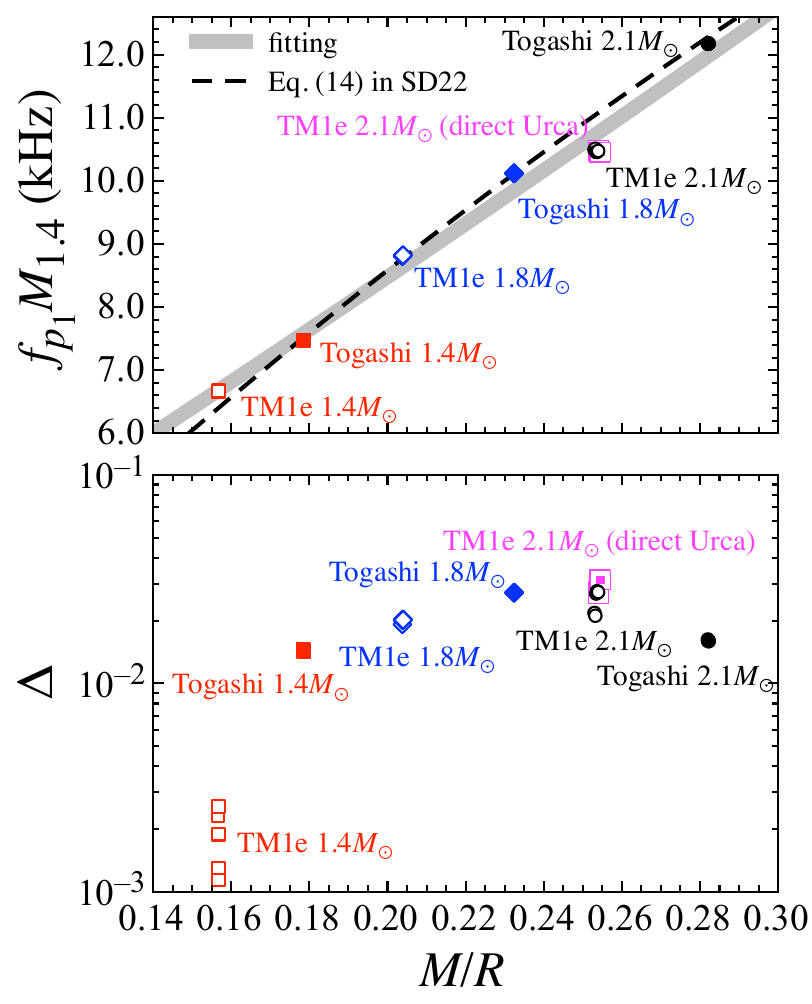} 
\end{center}
\caption{%%
Same as Fig.~\ref{fig:ffM}, but for the $p_1$-mode frequencies. The solid line is the fitting given by Eq.~(\ref{eq:fp1M}), while the dashed line denotes the fitting in the neutron star cooling derived in Ref.~\cite{SD22}.
}%%
\label{fig:fp1M}
\end{figure}

The dependence of the frequencies on the mass accretion rate in each stellar model is very weak, while the frequencies strongly depend on the stellar models, such as stellar mass and EOSs. To discuss such frequencies, the universal relation (or empirical relation) almost independent of the stellar models is important if any. For the $f$- and $p_1$-mode oscillations, we find that the frequencies can be well characterized with the fitting formulae given by
\begin{gather}
  f_f M_{1.4}\,{\rm (kHz)} = -0.8489 + 20.5416(M/R) -12.0398(M/R)^2,
  \label{eq:ffM} \\
  f_{p_1} M_{1.4}\,{\rm (kHz)} = 0.4790 + 38.1122(M/R) +9.3691(M/R)^2,
  \label{eq:fp1M} 
\end{gather}
where $M_{1.4}$ denotes the neutron star mass normalized with $1.4M_\odot$. In Figs.~\ref{fig:ffM} and \ref{fig:fp1M}, the frequencies multiplied with the stellar mass for various accreting neutron star models in the steady state are shown as a function of the stellar compactness, where the fitting lines given by Eqs.~(\ref{eq:ffM}) and (\ref{eq:fp1M}) are shown with the solid lines. To check how well these empirical relations work, we also show the relative deviation, $\Delta$, between the frequencies determined via eigenvalue problems and those estimated with the empirical relation in the bottom panels of each figure, i.e., $\Delta$ is calculated via
\begin{equation}
  \Delta = \frac{|f_{\rm fit} - f_{\rm Cow}|}{f_{\rm Cow}},
  \label{eq:Delta}
\end{equation}
where $f_{\rm fit}$ denotes the frequencies estimated with the fitting, such as Eqs.~(\ref{eq:ffM}) and (\ref{eq:fp1M}), while $f_{\rm Cow}$ denotes the frequencies determined with the eigenvalue problem. From these figures, we find that the $f$-mode ($p_1$-mode) frequencies are predicted with these empirical relations within less than $0.5\%$ (a few $\%$) accuracy. That is, through the observations of gravitational wave frequencies from the accreting neutron star in the steady state, one can get one constraint between the neutron star mass and radius, independently of the mass accretion rate and/or whether the direct Urca process works or not inside the star.

%%%%%%%%%%%%%%%%%%%%%%%%%%%%%%%%%%%
% Figure 6
%%%%%%%%%%%%%%%%%%%%%%%%%%%%%%%%%%%
%%%%%%%%%%%%%%%%%%%%%%%%%%%%%%%%%%%

%%%%%%%%%%%%%%%%%%%%%%%%%%%%%%%%%%%
% Figure 7
%%%%%%%%%%%%%%%%%%%%%%%%%%%%%%%%%%%
%%%%%%%%%%%%%%%%%%%%%%%%%%%%%%%%%%%

We have done a similar study on the cooling neutron stars~\cite{SD22}, where we could derive the empirical relations for the $f$- and $p_1$-mode frequencies, i.e., Eqs.~(13) and (14) in Ref.~\cite{SD22}. To compare them to the frequencies considered in this study, we also plot the empirical relations derived on the cooling neutron stars with the dashed lines in Figs.~\ref{fig:ffM} and \ref{fig:fp1M}. 
Using these empirical relations derived in Ref.~\cite{SD22}, we also calculate the relative deviation with Eq. (15), and then we find that it becomes at most $2\%$ in the $f$-mode and $6\%$ in the $p_1$-mode, where the $2.1 M_\odot$ models with TM1e with and without direct Urca become the worst.
From these figures, one can observe that the empirical relations in the cooling neutron stars are almost the same as in the accreting neutron stars in the steady state. Namely, the $f$- and $p_1$-mode frequencies are essentially determined with the stellar mass and radius, which are not so sensitive to the thermal profiles.

%%%%%%%%%%%%%%%%%%%%%%%%%%%%%%%%%%%
% Figure 8
%%%%%%%%%%%%%%%%%%%%%%%%%%%%%%%%%%%
\begin{figure}[tbp]
\begin{center}
\includegraphics[scale=0.5]{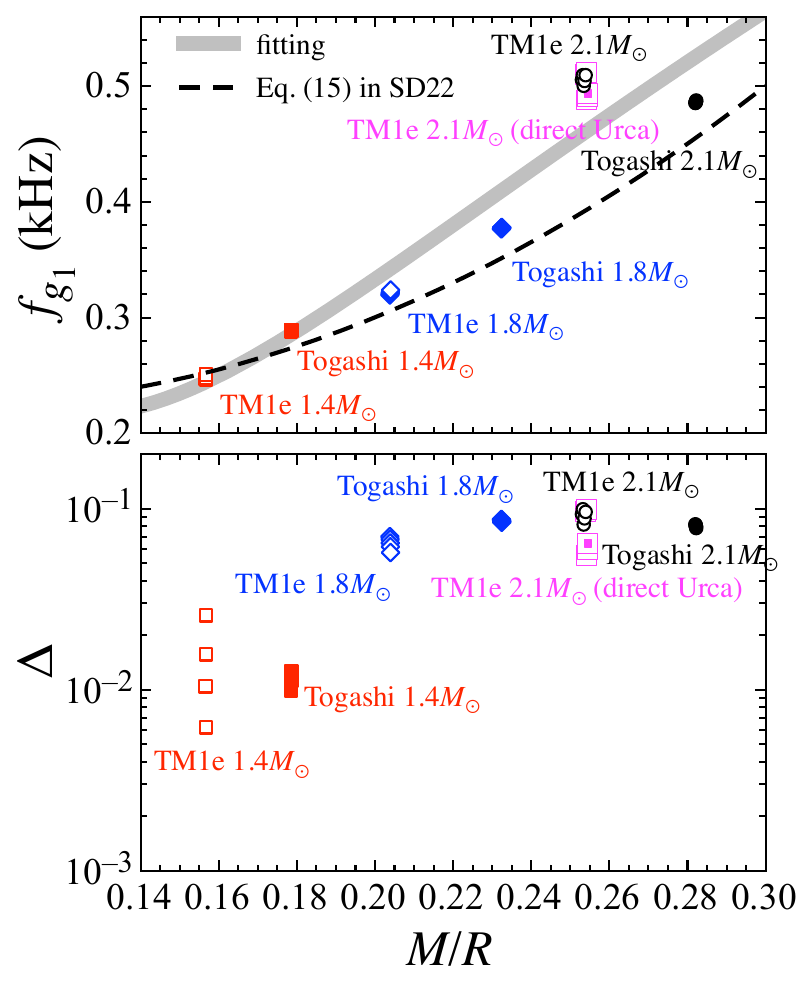} 
\end{center}
\caption{%%
The $g_1$-mode frequencies are shown as a function of $M/R$ for various accreting neutron star models in the steady state. The solid line denotes the fitting given by Eq.~(\ref{eq:fg1}), while the dashed line is the fitting in the neutron star cooling derived in Ref.~\cite{SD22}. The bottom panel denotes the relative deviation of the $g_1$-mode frequencies determined with the eigenvalue problem from the expectation with the fitting given by Eq.~(\ref{eq:fg1}), calculated with Eq.~(\ref{eq:Delta}).}%%
\label{fig:fg1MR}
\end{figure}
%%%%%%%%%%%%%%%%%%%%%%%%%%%%%%%%%%%

\begin{figure}[tbp]
\begin{center}
\includegraphics[scale=0.5]{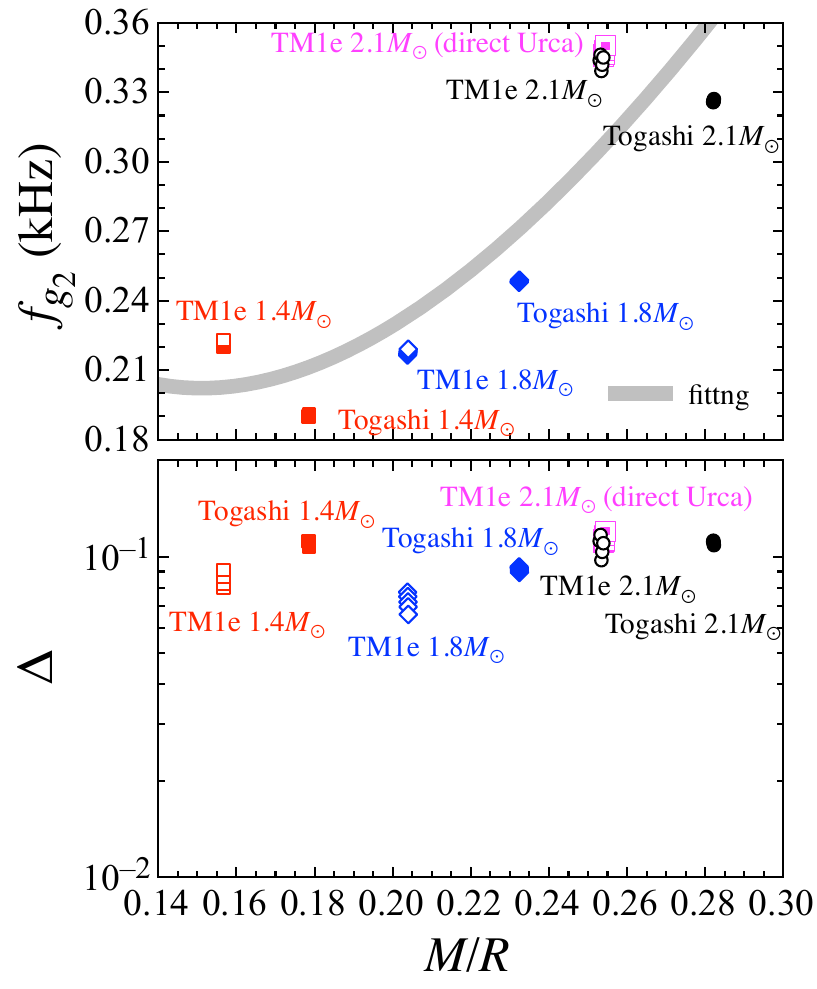} 
\end{center}
\caption{%%
Same as Fig.~\ref{fig:fg1MR}, but for the $g_2$-mode frequencies. The solid line is the fitting given by Eq.~(\ref{eq:fg2}).
}%%
\label{fig:fg2MR}
\end{figure}

On the other hand, we plot the $g_1$- and $g_2$-mode frequencies as a function of the stellar compactness in Figs.~\ref{fig:fg1MR} and \ref{fig:fg2MR}. Using these data, we derive fitting formulae, such as 
\begin{gather}
  f_{g_1}\,{\rm (kHz)} = 0.3708(R/M) -2.0715(R/M)^{1/2} +3.1111,
  \label{eq:fg1} \\
  f_{g_2}\,{\rm (kHz)} = 1.3938 -6.1322(M/R)^{1/2} +7.8893(M/R),
  \label{eq:fg2}  
\end{gather}
which are plotted with the solid line in the figures
\footnote{The functional form for the $g_2$-mode is different from that for the $g_1$-mode, because the $g_2$-mode frequencies from the $1.4M_\odot$ neutron star model constructed with TM1e seem to be different behavior. Maybe the $g_i$-mode frequencies for $i\ge 2$ depend more strongly on the thermal properties inside the star.}. Compared to the case with the $f$- and $p_1$-mode frequencies, the $g$-mode frequencies (especially the $g_2$-mode) deviate from the fitting line more. Even so, the $g_1$-mode ($g_2$-mode) frequencies are predicted with these empirical formulae within $10\%$ ($12\%$) accuracy. So, if one observes the $g_1$-mode (or $g_2$-mode) frequencies from the accreting neutron star in the steady state, one could get the constraint on $M/R$, which is another constraint on the neutron star mass and radius obtained from the observations of the $f$-mode (or $p_1$-mode) frequencies. That is, through the simultaneous observations of the $f$-mode (or $p_1$-mode) and the $g_1$-mode (or $g_2$-mode) frequencies, one could constrain the neutron star mass and radius, even if the mass accretion rate is unknown. 

%%%%%%%%%%%%%%%%%%%%%%%%%%%%%%%%%%%
% Figure 9
%%%%%%%%%%%%%%%%%%%%%%%%%%%%%%%%%%%
%%%%%%%%%%%%%%%%%%%%%%%%%%%%%%%%%%%

In the end, we have to mention an additional comment on the $g_1$-mode frequencies. In Fig.~\ref{fig:fg1MR}, we also plot the empirical relation obtained for the cooling neutron star with the dashed line \cite{SD22}, which seems to deviate more from the fitting line derived in this study. That is, unlike the $f$- and $p_1$-mode frequencies, the $g$-mode frequencies from the accreting neutron star in the steady state may be different from those for the cooling neutron star because the $g$-modes strongly depend on the thermal properties. Considering the temperature inside the neutron star focused on in this study, our results are consistent with those shown by Gittins and Andersson~\cite{GA24}, where the temperature effects of $g$-mode frequencies are discussed in isothermal neutron stars. 

%%%%%%%%%%%%%%%%%%%%%%%%%%%%%%%%%%%%%%%%%%%%%%%%
\section{Conclusion}
\label{sec:Conclusion}
%%%%%%%%%%%%%%%%%%%%%%%%%%%%%%%%%%%%%%%%%%%%%%%%

The accreting neutron star is potentially a continuous gravitational wave source. In this study, we examine the gravitational wave frequencies excited in the steady state of the accreting neutron stars. In such an object, the luminosity for an observer at infinity is one of the most important observables. In this study, we show that such a luminosity is well characterized by the mass accretion rate independent of the stellar mass and EOS if the direct Urca process does not work inside the star. On the other hand, the luminosity from the neutron star with direct Urca can deviate from such a characterization if the mass accretion rate is not so high. Thus, via the observations of luminosity, one could distinguish whether the direct Urca process works or not inside the star. In addition, we can derive the empirical relations for the gravitational wave frequencies for the $f$- and $p_1$-modes multiplied by the stellar mass as a function of the stellar compactness, which is independent of the mass accretion rate. Also, we can drive the other empirical relations for the $g_1$- and $g_2$-mode frequencies as a function of the stellar compactness. Thus, once one simultaneously observes the $f$-mode (or $p_1$-mode) and the $g_1$-mode (or $g_2$-mode) frequencies, one could get two constraints for the different combinations of the neutron star mass and radius, which enables us to constrain on the neutron star mass and radius.

%\newpage
%%%%%%%%%%%%%%%%%%%%%%%%%%%%%%%%%%%%%%%%%%%%%%%%
\acknowledgments
%%%%%%%%%%%%%%%%%%%%%%%%%%%%%%%%%%%%%%%%%%%%%%%%

This work is supported in part by Japan Society for the Promotion of Science (JSPS) KAKENHI Grant Numbers 
JP21H01088,  % Kiban(B) by Sotani
JP23K20848,  % Kiban(B) by Sotani
JP23K19056,  % Start-up by Dohi
and JP24KF0090, % by Sotani & Kumar
and by FY2023 RIKEN Incentive Research Project.

\appendix
%%%%%%%%%%%%%%%%%%%%%%%%%%%%%%%%%%%%%%%%%%%%%%%%
\section{Behavior of eigenfunctions}   % Appendix A
\label{sec:appendix_1}
%%%%%%%%%%%%%%%%%%%%%%%%%%%%%%%%%%%%%%%%%%%%%%%%

\begin{figure}[tbp]
\begin{center}
\includegraphics[scale=0.6]{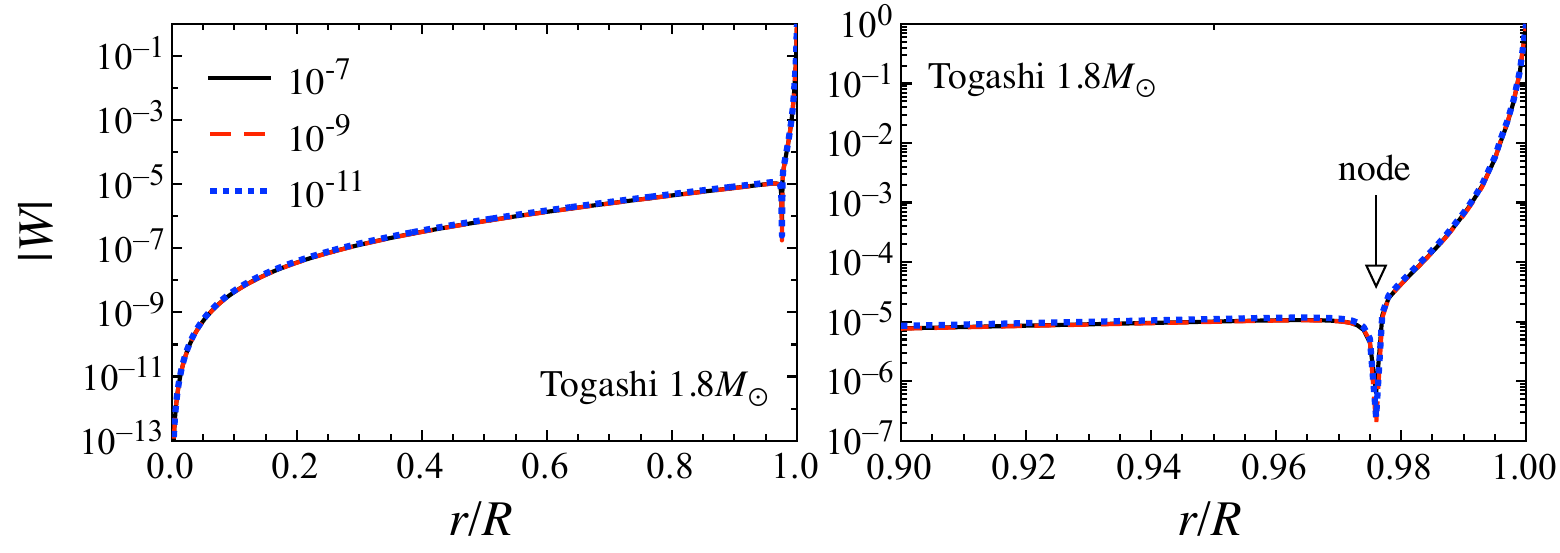} 
\end{center}
\caption{%%
For the ``f-mode" frequency ($2.54$ kHz), the radial profile of the absolute value of the eigenfunction, $W$, which is the radial displacement of the fluid element, for the $1.8M_\odot$ model with Togashi. The solid, dashed, and dotted lines correspond to the results for $\dot{M}=10^{-7}$, $10^{-9}$, and $10^{-11}$ $M_\odot$/yr, respectively. The right panel is an enlarged view of the left panel. 
}%%
\label{fig:WrTG18}
\end{figure}

The eigenmodes are simply classified by counting the nodal number of the corresponding eigenfunctions. In the usual cases, such as cold neutron stars, the eigenfrequency with which the nodal number is zero, is identified as the $f$-mode, while the eigenfrequencies with which the nodal number is $i$, are identified as the $g_i$-modes ($p_i$-modes) if they are lower  (higher) than the $f$-mode frequencies. On the other hand, since at least the $f$ and $p$-modes strongly depend on the stellar mass and radius, especially on the stellar average density, as mentioned in the text,  the eigenfrequencies in the zero-accretion limit should become similar values for the cold neutron stars with the same mass and radius. Even so, we find that the behavior of the eigenfunction is a little strange in the vicinity of the surface. For example, regarding the ``f-mode" frequency in the accreting neutron star, the radial profile of the eigenfunction, e.g., the radial displacement of the fluid element ($W$), is plotted in Fig.~\ref{fig:WrTG18}. By definition, the $f$-mode has no node, but as shown in this figure this mode has one node. Unfortunately, we did not identify the origin of this strange behavior, but one possibility is the conventional instability, where the Brunt-V\"{a}is\"{a}l\"{a} frequency, $f_{\rm BV}$, becomes negative. As shown in Fig.~\ref{fig:BV}, $f_{\rm BV}$ becomes strongly negative in the vicinity of the stellar surface, which indicates that such a region is strongly unstable. Since this behavior does not appear in the cooling neutron star, e.g., \cite{SD22}, this seems to be a specific feature in the accreting neutron star. Anyway, as mentioned in the text, we consider that the eigenfrequencies, which are almost the same as those for the cold neutron star with the same mass, are identified as the corresponding eigenmodes in this study.

%%%%%%%%%%%%%%%%%%%%%%%%%%%%%%%%%%%
% Figure 10
%%%%%%%%%%%%%%%%%%%%%%%%%%%%%%%%%%%
%%%%%%%%%%%%%%%%%%%%%%%%%%%%%%%%%%%

%\bibliographystyle{h-physrev} % for PrD
%\bibliography{mybib}
%%%%%%%%%%%%%%%%%%%%%%%%%%%%%%%%%%%%%%%%%%%%%%%%

\end{document}